\documentclass[12pt,preprint]{aastex}

\shorttitle{New nearby M dwarfs}
\shortauthors{Riaz, Gizis & Harvin}

\begin{document}

\title{Identification of new M dwarfs in the solar neighborhood}

\author{Basmah Riaz, John E. Gizis and James Harvin}
\affil{Department of Physics and Astronomy, University of Delaware,
    Newark, DE 19716; basmah@udel.edu, gizis@udel.edu, harvin@udel.edu}

\begin{abstract}

We present the results from a spectroscopic study of 1080 nearby active M dwarfs, selected by correlating the 2MASS and ROSAT catalogs. We have derived the spectral types and estimated distances for all of our stars. The spectral types range between K5 and M6. Nearly half of our stars lie within 50 pc. We have measured the equivalent width of the H$\alpha$ emission line.  Our targets show an increase in chromospheric activity from early to mid-spectral types, with a peak in activity around M5. Using the count rate and hardness ratios obtained from the ROSAT catalog, we have derived the X-ray luminosities. Our stars display a ``saturation-type'' relation between the chromospheric and coronal activity. The relation is such that log $L_{X}$ $\slash$ $L_{bol}$ remains ``saturated'' at a value of $\sim$ -3 for varying H$\alpha$ equivalent width. We have found 568 matches in the USNO-B catalog, and have derived the tangential velocities for these stars. There is a slight trend of decreasing chromospheric activity with age, such that the stars with higher $v_{tan}$ have lower H$\alpha$ equivalent widths. The coronal emission, however, remains saturated at a value of log $L_{X}$ $\slash$ $L_{bol}$ $\sim$ -3 for varying tangential velocities, suggesting that the coronal activity remains saturated with age. We do not find any break in the saturation-type relation at the spectral type where stars become fully convective ($\sim$M3.5). Most of the stars in our sample show more coronal emission than the dMe stars in the Hyades and Praesepe clusters, and have $v_{tan}$ $<$ 40km/s, suggesting a young population.

\end{abstract}


\section{Introduction}

In order to fully understand the atmospheric and kinematics properties of the stars, a detailed study of the most numerous stars in the Galaxy, the M dwarfs, is essential. The spectra of these late-type main sequence stars are noticeably affected by their magnetic activity. The non-radiative, magnetic field responsible for heating the chromospheres and coronae of these low-mass stars is generated by a dynamo. For Sun-like stars, the shell dynamo is the principal mechanism responsible for the generation of the magnetic flux. Whereas, in fully convective stars like late-type M dwarfs, the shell dynamo fails due to the absence of a radiative core. The turbulent dynamo (Durney et al. 1993), however, appears to explain the stellar physics in such stars well since it does not require a radiative- convective boundary layer and can produce magnetic fields by random convection motions in the convection zone (Reid \& Hawley, 2000). Magnetic fields in the range 2-4 kG have been measured on M dwarfs, which are twice as strong as the fields measured on G and K stars (Johns-Krull \& Valenti, 1996). Located in the red at 6563 $\AA$, the H$\alpha$ line is the best diagnostic of an active chromosphere in an M dwarf, in comparison with the Ca II H and K lines which are less accessible, and X-ray emission is indicative of the presence of a corona. 

This paper is an effort to study the magnetic activity in field M dwarfs, and to investigate if there exists any relation between their chromospheric and coronal activity.  Previous attempts to investigate this problem have concentrated on star clusters, or on low-mass stars in individual catalogs. Fleming et al. (1993) conducted a ROSAT survey of the late M dwarfs, and found that these stars can have coronae as active as early M dwarfs. Later, Schmitt, Fleming \& Giampapa (1995) studied the coronal activity of all known K and M type stars within 7 pc of the Sun using X-ray data from the ROSAT All-Sky Survey (RASS). They obtained a detection rate of 87\%, implying that most K and M dwarfs have active coronae. Fleming et al. (1995) found that the dMe stars that exhibit the greatest amount of chromospheric activity also exhibit the greatest levels of coronal activity. Hodgkin et al. (1995) conducted a study of the chromospheric and coronal activity of a large sample of low mass and very low mass stars in the Pleiades. They found that chromospheric activity is suppressed with respect to coronal activity for stars with $M_{I}$ $>$ 8. Barrado et al. (1998) examined the activity in the Praesepe and Hyades clusters. They found a lower X-ray detection rate in the Praesepe than in the Hyades, resulting in different luminosity functions. Also, a strong correlation was found between chromospheric and coronal activity in the Hyades cluster.  

Hawley et al. (1996) showed that the fraction of active M dwarfs increases from early to mid-spectral types, and reaches a peak around M7. Gizis et al. (2000) added the cooler dwarfs and found a decline in activity at spectral types later than M7. Fleming et al. (1993) showed that there is no decrease in the X-ray flux, and thus, the coronal heating efficiency among the late M dwarfs, and that if there is any sharp decline at all, then it occurs at spectral type M8. This suggests an increase in both chromospheric and coronal activity from early to mid-type M dwarfs. We have conducted a spectroscopic survey of candidate M dwarfs selected by correlating the 2MASS and ROSAT X-ray catalogs. We have used H$\alpha$ emission and X-ray fluxes as indicators of chromospheric and coronal activity, respectively, to study the behavior of magnetic fields in our sample of stars. Our entire sample, being X-ray selected, is biased towards coronally active stars. Coronal X-ray emission provides a tool by which unknown nearby stars can be discovered. Thus our study has helped in identifying new nearby active M dwarfs.

\section{Instrumentation}

We have obtained medium-resolution infrared spectra of the stars in our sample using the Palomar 60-inch telescope for the northern stars, and the CTIO 1.5m telescope for the southern hemisphere ones. Data from the Palomar 60-inch was obtained in October, November, 2000, and in June, 2001. The detector used was an 800 $\times$ 800 format TI CCD. The grating used had 600 lines$\slash$mm, with a blaze wavelength of 6500 $\AA$, and wavelength coverage of 6200 to 7400 $\AA$. The dispersion was 1.5$\AA$$\slash$pix. The slit width was set to 1$\arcsec$, giving a resolution of $\sim$ 2.5 pixels.

Spectra for the southern stars were obtained during a series of observing runs from September, 2002 to May, 2004. The instrument used was the 1.5-meter Cassegrain spectrograph. This spectrograph uses a semi-solid Schmidt camera with 160 mm focal length. A Loral 1200 $\times$ 800 CCD is used with the camera, and has a pixel size of 15 microns. The gain was set to 4, which corresponds to a read time of 34 seconds. The slit width was set to 1.5$\arcsec$. Our wavelength range of interest was 6200 to 7600 $\AA$. For this purpose, we chose the OG 570 filter. It is a blue-blocking long-pass filter, and blocks all light short ward of approximately 5900 $\AA$. The quantum efficiency of the Loral 1K in this wavelength range is between 91-93\%. The grating used was 35(I), with a tilt of 19.64$\degr$. It has 600 lines$\slash$mm, with a first-order resolution of 4.3 $\AA$. The blaze wavelength is 6750 $\AA$. A neon discharge tube was used as a comparison source

\section{Sample Selection}

M dwarfs being cool stars, emit most of their energies in the IR band.  However, these late-type stars also display coronal activity in the form of X-ray emission. We started with choosing stars from the 2MASS catalog with the requirement that the galactic latitude be greater than 15 or less than -15, in order to avoid the high density of stars at very low latitudes.  Then, color cuts were applied that corresponded to (J-H) $<$ 0.75, (H-K) $>$ 0.15 and 0.8 $<$ (J-K) $<$ 1.1. This provided us with stars that had J, H, and K magnitudes between 5 and 12, similar to the M dwarf colors. We then matched these to the ROSAT X-ray sources, with a matching radius of 20$\arcsec$. Our final sample consisted of 2403 stars; out of these, we have taken spectra of 352 northern hemisphere stars, and 728 southern hemisphere ones. Figure 1 shows a plot of H-K vs. J-H for our sample. Table 1 lists their observed and derived quantities. Some of these have been identified by previous surveys (Carpenter et al. 2001, Craig et al. 1997, Fuhrmeister et al. 2003, Koehler 2001, Li et al. 1998, Li et al. 2000, MacConnell 1981, Massey 2002, Preibisch et al. 1999, Thompson et al. 1998, Zickgraf et al. 2003). In principle, some giants could be included in our sample, but as discussed in Section ~\ref{spty}, the spectra for none of these classify them as giants.

\section{Observations}

During each night, we took ten bias, dome flat and comparison calibration frames. Additionally, spectra of some flux standards were taken for flux calibration purposes. The IRAF CCD reduction package, ccdred, was used for the subtraction of an overscan bias, subtraction of a zero level image, and the division by a flat field calibration image, for all of the object spectra. The overscan corrected and flat fielded spectra were then extracted using the IRAF task, apall. The extracted spectra were then wavelength-calibrated using the dispersion solution determined for the comparison spectra. Finally, the object spectra were flux-calibrated using the flux standard spectra.

\section{Discussion}

\subsection{Molecular Band Indices and Spectral Types}
\label{spty}

The spectra of M dwarfs are dominated by molecular bands of TiO (bandheads at 6500 and 7050 $\AA$), CaH ($\sim$6400 and $\sim$6800 $\AA$) and CaOH (6250$\AA$), and the atomic lines of Ca II, Na I and K I. There are also telluric lines of $H_{2}O$ and the telluric A band and B band of $O_{2}$. The absorption in the CaH bands increases with decreasing metallicity, which is why these are useful in classifying the halo subdwarfs and the metal-poor disk dwarfs. The absorption in the TiO bands increases from early to mid-M types. The spectra of the latest M dwarfs (M7 and later) show weak TiO absorption, but prominent absorption by vanadium oxide (Kirkpatrick et al. 1991, Reid et al. 1995).

We have measured the bandstrengths for different bands using the IRAF task, sbands. The bandpasses selected were the same as defined in Reid et al. (1995).  Since in the input bandpass file, two bandpasses were entered for each spectral feature of interest, the output file consisted of the fluxes for each of the two bandpasses, as well as a band index, or the flux ratio. We have multiple observations of some 20 stars. Using these, we estimate the uncertainty in our measurement of the indices to be $\pm$0.01-0.03.

Using the spectral slopes and strengths of the red/near-infrared spectral features, Kirkpatrick et al. (1991) have built a classification system based on the spectra of K and M dwarf standards, over the wavelength range 6300-9000 $\AA$. All spectral features contribute to some extent to the final spectral type (SpT) determination. Reid et al. (1995) have used a specific feature (the full depth of the strongest TiO feature at 7050$\AA$) as the primary SpT indicator, and have defined their own system, by tying their observations to the Kirkpatrick et al. (1991) system through using their SpT standard stars. We have used the relation from Reid et al. (1995) to obtain SpT for our stars:	

\begin{equation}
SpT = -10.775 ~(TiO5) + 8.2 
\end{equation}

\noindent Reid et al. (1995) report an uncertainty of $\pm$0.5 subclass. The SpT of our stars range between K5 to M6. Since this relation is valid only in the spectral range $\sim$K7 to M6$\slash$M6.5, for stars earlier than K7, we have confirmed their SpTs by comparing with the spectral grid from Allen \& Strom (1995). Their resolution is similar to ours ($\sim$4 $\AA$). None of our stars can be classified as giants. Giants display weaker absorption in the 6750-7050 $\AA$ CaH band than dwarfs, as discussed in Allen \& Strom (1995). Comparing
with their M giant spectra, we have found that our spectra show
more absorption and thus must be of higher surface gravity. Also, comparing with Fig. 4 in Reid et al. (1995), giants appear as outliers and lie above the main body of dwarf stars in the CaH3 (6750-7050 $\AA$ CaH band) vs. TiO5 (bandhead at 7050$\AA$) indices diagram. The absence of any such outliers for our sample confirms that our targets are all field dwarf stars.

\subsection{Absolute Magnitudes and Distances}

Fig. 2 shows a plot of $M_{J}$ vs. the TiO5 indices for the 44 stars in our sample with known trigonometric parallaxes. Filled triangles represent stars for which the parallax error is less than 10\%, while the ones with errors greater than 10\% are denoted by open triangles. There are two stars for which the parallax error is not known, and these are denoted by crosses. We can see a break at TiO5$\sim$0.4. For stars with TiO5$\ga$0.4, the best-fit relation is a second-order polynomial given by

\begin{equation}
M_{J} = 0.97~ (TiO5)^2 - 5.01 ~(TiO5) + 8.73,
\end{equation}

\noindent with an rms scatter of $\pm$0.8 mag. For stars with TiO5$<$0.4, we have used the relation from Cruz \& Reid (2002), given as

\begin{equation}	
M_{J} = -7.43~ (TiO5) + 11.82,
\end{equation}

\noindent with an uncertainty of 0.19 mag. These two relations are represented by thick solid lines in Fig. 2. Stars that lie in the overlapping region around TiO5$\sim$0.4 have a high uncertainty in their absolute magnitudes and the derived distances, since these could be brighter or fainter than the $M_{J}$ obtained from the above relations.

The thin solid line in Fig. 2 represents the absolute magnitude calibration relation used by Cruz \& Reid (2002) for TiO5$\ga$0.34. As can be seen, this relation suggests $M_{J}$ that are fainter by about a magnitude, as compared to the ones derived from relation (2) above. Hawley et al. (1996) found that the dMe stars are brighter than their dM counterparts for SpT $\la$ M3, whereas, they fit the mean dM relation well at later types. Although relation (2) is a fit to very few data points, almost all of our targets are dMe stars, which suggests brighter absolute magnitudes.

The bolometric magnitudes and luminosities were derived from the absolute K magnitudes using the relation from Veeder (1974):  	

\begin{equation}
M_{bol} = (1.12) ~M_{K} + 1.81 
\end{equation}

\noindent Veeder (1974) give a scatter about this mean relation of 0.15 mag. Using the bolometric corrections from Leggett et al. (2000) gave similar values. 

We have estimated the distances for our stars using the absolute magnitudes derived from relations (2) and (3). The uncertainties in the distances are on the order of $\pm$37\%. Nearly half of our stars lie within 50 pc. There are 41 stars that lie within 10 pc. We have the distances measured from parallax for 12 of these, three of which are greater than 10 pc.

\subsection{Chromospheric Activity}

Magnetic heating of the chromosphere gives rise to the most accessible indicator of activity, the H$\alpha$ emission line, in M dwarfs. We have measured the equivalent width (EW) of this line using the IRAF task, splot. The uncertainty in EW is dominated by the uncertainty in defining the continuum level, and is of the order of 0.5$\AA$. Figs. 3 and 4 show plots of the H$\alpha$ EW vs. the TiO5 index. We find a trend of increasing EW as we move towards cooler M dwarfs, with a peak in activity around TiO5 $\sim$ 0.3 (SpT $\sim$ M5), implying cooler, lower mass stars to be more active than the hotter ones. Stars of later spectral types retain their activity for a longer period of time, due to longer spin-down timescales. The spin-down timescale is of the order of a few Gyr for SpT M3-M4, and of the order of 10 Gyr at SpT M6 (Delfosse et al. 1998). Thus we see an increase in the chromospheric activity from early to mid-spectral types. There is a little tail at TiO5 $\sim$ 0.9. The H$\alpha$ line is in absorption for these stars, showing a lack of chromospheric activity. The lack of TiO5 absorption suggests a poor estimate of their SpT. 

In Fig. 3, open circle represents the T Tauri-type star TW Hya that shows very strong H$\alpha$ in emission. It is known to be a strong accretor. We have observed this star twice, once in June 2003 and then in February 2004. It showed high activity both times. Stars denoted by crosses display the strongest emission in H$\alpha$ among the early type M dwarfs, while the one star denoted by a triangle shows the strongest emission among the later types. Table 2 lists their names and other parameters, along with other stars that display strong emission in the H$\alpha$ line (EW$>$ 20$\AA$). In Fig. 4, the stars that have H$\alpha$ EW greater than 8$\AA$, but show weak TiO5 absorption are denoted by squares. Three of these are T Tauri-type stars. Their names are also listed in Table 2

\subsection{Coronal Activity}

In order to determine the X-ray fluxes for our targets, we have used the count rate-to-energy flux conversion (CF) relation found by Schmitt, Fleming and Giampapa (1995). The relation is:             	

\begin{equation}
CF = (5.30 ~HR + 8.31)~ 10^{-12}  ~~~~ergs \slash cm^2 \slash counts
\end{equation}

\noindent Then, the X-ray flux was calculated as:    
                       
\begin{equation}
F_{X} = CF \times CR  
\end{equation}

\noindent The hardness ratios (HR) and count rates (CR) were obtained from ROSAT. Using the X-ray fluxes and the distances, we have calculated the X-ray luminosities. All of our targets have $L_{X}$ in the range $10^{27}$ - $10^{33}$ ergs$\slash$sec. Barrado, Stauffer and Randich (1998), found log $L_{X}$ values for the strongest X-ray emitters among the dMe stars in the Praesepe and Hyades clusters to be $\sim$29-29.5. Most of the stars in our sample show more coronal emission and thus should be younger than the dMe stars in these clusters. 

Fig. 5 shows a plot of log $L_{X}$$\slash$$L_{bol}$ vs. SpT. The X-ray luminosities remain ``saturated'' with varying SpT. We can also see a large spread in $L_{X}$ for a given SpT. As suggested by Stauffer et al. (1994, 1997), the dispersion in $L_{X}$ for low mass stars in young clusters like Pleiades and Hyades can be explained by the rotational velocity differences. The large range in rotational velocities is due to the spread in the mass and lifetimes of the circumstellar disks around low mass stars; stars with short-lived disks become rapid rotators, while the ones with long-lived disks become slow rotators (Stauffer et al 1997). This spread in $L_{X}$ has also been seen in other clusters such as the Hyades (Stern et al. 1995), IC 2391 (Patten \& Simon 1993), the Praesepe (Randich \& Schmitt 1995) and IC 2602 (Randich et al. 1995). Thus the few showing extraordinarily high activity in our sample (log $L_{X}$$\slash$$L_{bol}$ $\sim$ -2) might be very rapid rotators. The large dispersion in the X-ray luminosities seen for the late-K stars decreases towards early M-types, as can be seen from Fig. 5.  Stauffer et al. (1994) explain this to be caused by the longer spin down timescales for the redder stars. For the field M dwarfs, the spin-down timescale is a few Gyr at spectral type M3-M4 (Delfosse et al. 1998). Due to the longer spin down timescales, the redder stars have rotational velocities higher than the threshold (v sin i $>$ 15km/s), which results in an increase in the fraction of stars in the saturated portion, thus decreasing the dispersion in $L_{X}$ for the late-type dwarfs (Stauffer et al. 1997).

We did not find any break in log $L_{X}$$\slash$$L_{bol}$ at SpT$\sim$M3.5, where the stellar interior changes from a radiative core, convective envelope to being fully convective. Delfosse et al. (1998) did not find any break in the rotational velocity distribution and the rotation/activity relations for field M dwarfs at the spectral type where stars become fully convective, and suggested that the turbulent dynamo already drives most of the magnetic field in stars that still have a radiative core. The sharp decline in X-ray luminosity seen by Fleming et al. (1993) occurs at spectral type $\sim$M8, which implies masses lower than the critical mass, and is thus not directly related to the transition to a fully convective stellar interior. Indeed, these authors have also concluded that coronal heating efficiencies do not decrease toward the totally convective stars near the end of the main sequence.

To understand the relation between coronal and chromospheric activity, we have plotted log $L_{X}$$\slash$$L_{bol}$ vs. H$\alpha$ EW in Fig. 6. As can be seen, log $L_{X}$$\slash$$L_{bol}$ remains mainly ``saturated'' at a value of $\sim$ -3, with varying H$\alpha$ EW. Stauffer et al. (1994) found that the X-ray luminosities for low-mass stars in the Pleiades rise rapidly with increasing rotation rates, and then reach a ``saturated''  value of log $L_{X}$$\slash$$L_{bol}$ $\sim$ -3, for vsin i $>$15 km$\slash$s. They thus concluded that rotation is the dominant determinant of X-ray luminosity, and that this ``turn-on''  of rotationally driven dynamo activity occurs around (B-V) $\sim$ 0.6. A similar ``saturated-type''  rotation/activity relation was seen for the low-mass stars in the Hyades cluster (Stauffer et al. 1997). Delfosse et al. (1998) also found a saturation-type rotation/activity relation for field M dwarfs. They found a lower threshold for v sin i $\sim$4-5 km$\slash$s, similar to the one found for low-mass stars with long-lived circumstellar disks in the Hyades (Stauffer et al. 1997). As pointed out by Peterson (1989), the volume of the convection zone in fully convective stars decreases in proportion to the mass, which affects the magnetic field generation. The field strength may become saturated in the lowest mass stars, which would explain the absence of a strong rotation/activity relation (e.g., Stauffer et al. 1991). Our sample of field M dwarfs suggests a range in ages, as compared to individual clusters, where all members are assumed to be at the same age. If age was the dominant factor, we would have seen some change in activity for the older stars in our sample. The fact that we don't suggests that the dynamo that generates the magnetic field responsible for heating the chromospheres and coronae in low-mass stars is rotationally driven. However, we note that our stars were selected based on their strong X-ray emission. Thus all of them are expected to lie in the ``saturated'' region, independent of their age. In such an active sample, we do not expect to see a turn-off of activity for the older stars.

Fleming et al. (1995) found a correlation between chromospheric and coronal activity for M dwarfs within 7 pc of the Sun. We do not see such a correlation for our stars within this distance limit. The saturation type relation seen for stars at larger distances is seen for the nearby ones too.

Giampapa et al. (1996) found the coronal heating efficiency to be 2-3 orders of magnitude smaller for dM stars than for active dMe stars. We have found log $L_{X}$ $\sim$ 26-32 and log $L_{X}$$\slash$$L_{bol}$ $\sim$ -2 - -4 for both dM and dMe stars in our sample. This large spread is due to the differences in rotational velocities, as discussed earlier in this section.

\subsection{Proper Motions}

We have obtained proper motions for our targets by matching them with the USNO-B catalog (Monet et al. 2003). Out of our 1080 targets, we found matches for 568 stars. Using the proper motions and the distances, we were able to determine the tangential velocities for our stars. The errors in the proper motions are small (0.001-0.1$\arcsec$/yr). But due to the large uncertainty in the distance measurements ($\sim$37\%), we estimate similar uncertainties for the $v_{tan}$ values. There are four stars with $v_{tan}$ $>$100. Their names and other parameters are listed in Table 3. Their proper motions are not very high (0.2-0.6$\arcsec$/yr), but their large distances could result in large tangential velocities. Figure 7 shows a plot of H$\alpha$ EW vs. the tangential velocities. We find a large spread in $v_{tan}$ for a given value of H$\alpha$. Most of the stars in our sample have $v_{tan}$ $<$40 km$\slash$s, implying that they are a young population. From Fig. 7, we find that the targets that show very strong H$\alpha$ emission have small tangential velocities, whereas the ones with very high $v_{tan}$ show low activity levels. Thus there is a slight trend of decreasing chromospheric activity with age. Fig. 8 shows $v_{tan}$ vs. log $L_{X}$$\slash$$L_{bol}$. The X-ray luminosity remains mostly saturated around the mean value of log $L_{X}$$\slash$$L_{bol}$ $\sim$ -3 for varying tangential velocities, suggesting that the coronal activity remains saturated with age. This again suggests that rotation, and not age, dominates the magnetic activity in low-mass stars.

\section{Targets of Interest}

\subsection{Within 5pc}

We have found five stars that lie within 5 pc, and have confirmed the distances for three of these using parallaxes. Four of these are previously known stars. Table 4 lists their names, spectral types and distances. The distances derived from parallaxes are also given. One of these, 2MASS J06045215-3433360, was reported to lie at a distance of 6.1 $\pm$ 1.3 pc (Scholz et al. 2005). Since the parallax for this star is not known, we are not certain about the distance we have determined, due to the high uncertainty in the distance measurements.

\subsection{Strong X-ray Emitters}

We have found 32 stars that show strong X-ray emission (log $L_{X}$$\slash$$L_{bol}$ $>$ -2), which makes them interesting T Tauri-type candidates for future studies. Most of these are late K dwarf stars that lie at d$>$100pc and have H$\alpha$ in absorption. The 2MASS images for 19 of these show that they lie in confused regions of the sky, which could explain their strong X-ray emission. One star (2MASS J20194981-5816431), in particular, shows strong emission in H$\alpha$ and has log $L_{X}$$\slash$$L_{bol}$ = -2.3. We have observed this star twice and it displays variability in H$\alpha$ line strength (EW=45.4 $\AA$ and 39.3 $\AA$). It is a M6 star and lies at a distance of 12pc from the Sun.

\section{Summary}

We have identified new nearby M dwarfs, by correlating the 2MASS and ROSAT catalogs. The spectral types derived for these stars range between K5-M6. Almost all lie within 100pc of the Sun, with nearly half lying within 50 pc. The chromospheric activity of these stars rises towards later spectral types, with a peak around M5. A saturation-type relation is seen between chromospheric and coronal activity. The relation is such that log $L_{X}$$\slash$$L_{bol}$ remains saturated at $\sim$ -3 with increasing H$\alpha$ EW. While we see a slight trend of decreasing chromospheric activity with age, the X-ray luminosity remains saturated for varying tangential velocities, suggesting that the coronal activity remains saturated with age. The saturation value is similar to the one seen for rotation/activity saturation-type relations in the Hyades and Pleiades clusters, based on which Stauffer et al. (1994, 1997) concluded that the magnetic dynamo is rotationally driven. Our sample of field stars suggests a wide range in ages. Since we do not see any rough correlation of X-ray emission with age suggests that rotation, not age, dominates the magnetic activity in low-mass stars. We do not find any break in the X-ray emission at the spectral type where stars become fully convective; X-ray emission remains saturated with increasing spectral types. There is also a large spread in X-ray luminosities for a given spectral type, which reflects the range in the rotational velocities of the stars. We have found five stars that lie within 5 pc, and have confirmed the distances for three of these using parallaxes. There are 32 stars that show strong X-ray emission (log $L_{X}$$\slash$$L_{bol}$ $>$ -2), which makes them interesting T Tauri-type candidates for future studies.

\acknowledgments
This work has made use of the SIMBAD database, the 2MASS NIR, ROSAT All Sky Survey, and the USNO-B catalogs. B.R. would like to thank Dr. Steve Strom for his useful suggestions. This work has been partially supported by a grant from the University of Delaware Research Foundation.

\clearpage

\begin{figure}
\resizebox{150mm}{!}{\includegraphics[angle=270]{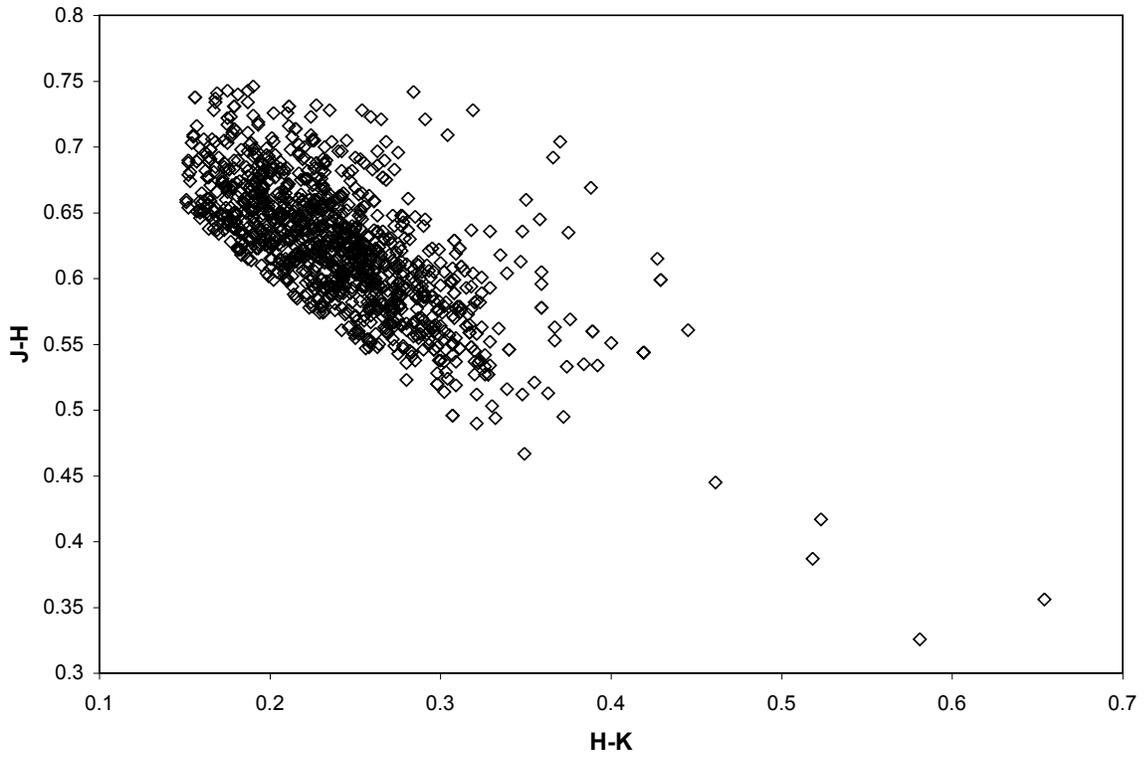}}
\caption{J-H vs. H-K for our sample.}
\end{figure}

\clearpage

\begin{figure}
\resizebox{150mm}{!}{\includegraphics[angle=270]{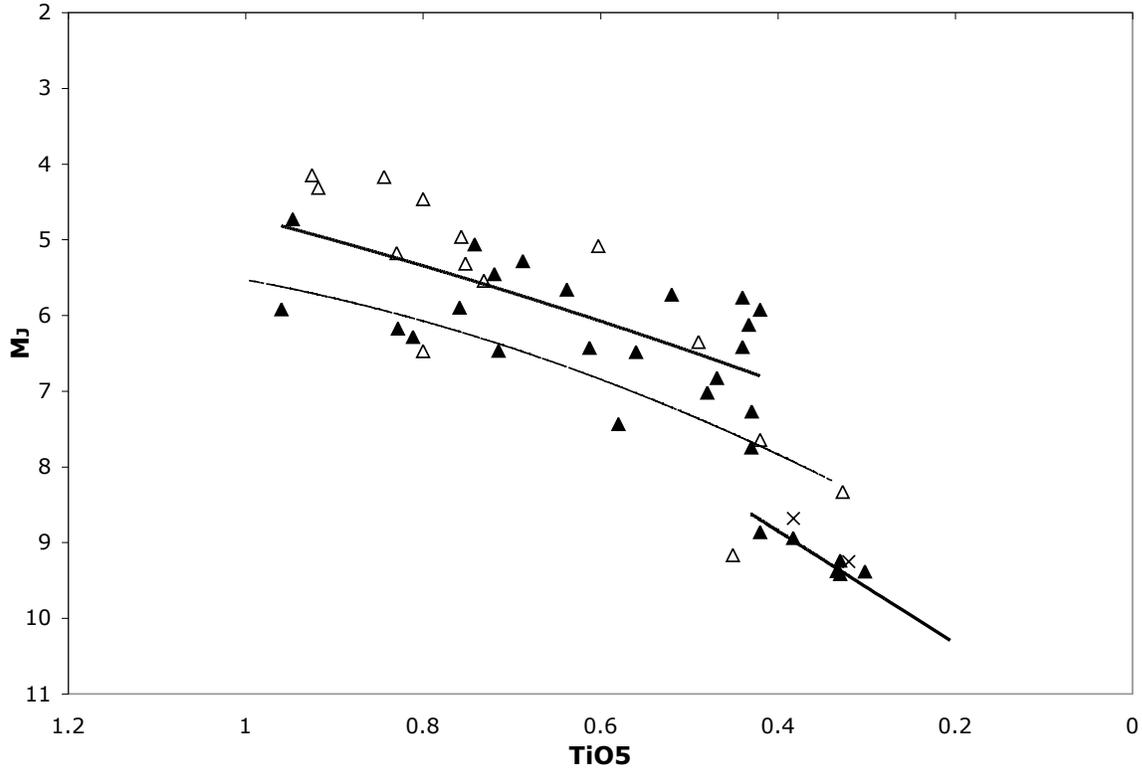}}
\caption{Derived $M_{J}$ vs. TiO5 indices. Symbols are as follows: Filled triangles -- stars with parallax error less than 10\%, open triangles -- stars with parallax error greater than 10\%. Thick solid lines are the best-fit to the absolute magnitudes derived from parallaxes. Thin solid line is the relation from Cruz \& Reid (2002).}
\end{figure}

\clearpage

\begin{figure}
\resizebox{150mm}{!}{\includegraphics[angle=270]{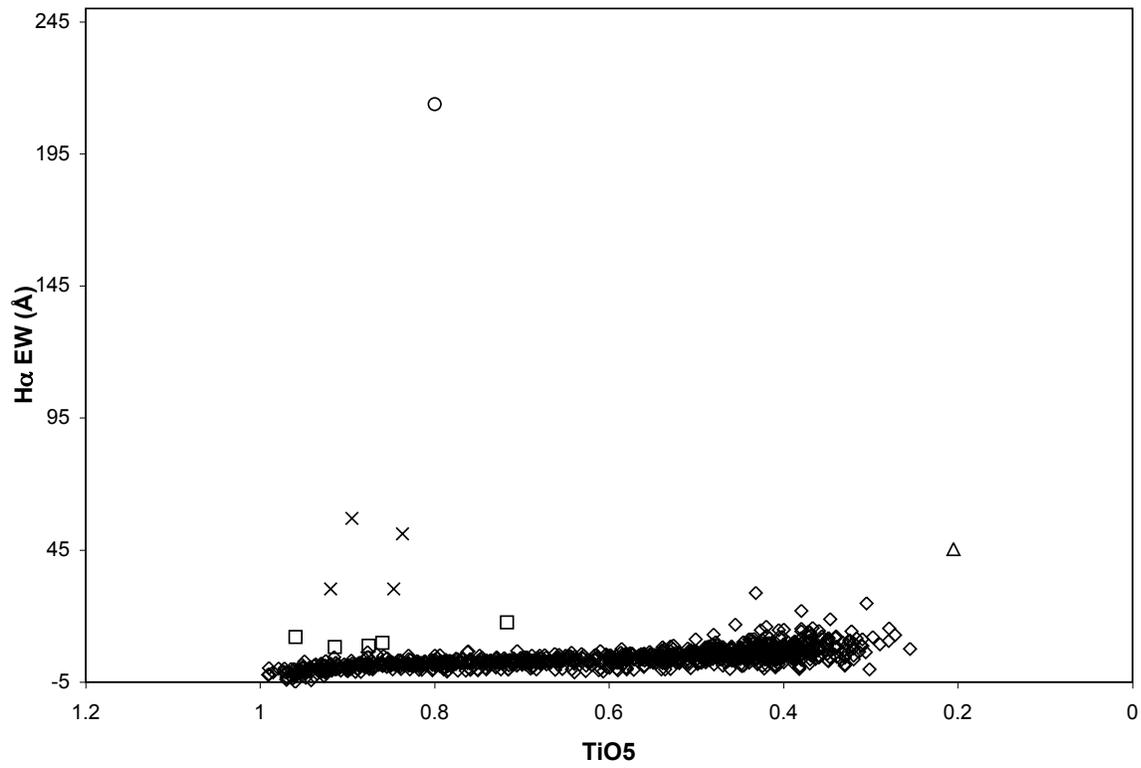}}
\caption{H$\alpha$ EW vs. the TiO5 indices. Chromospheric activity rises towards later spectral types. }
\end{figure}

\clearpage

\begin{figure}
\resizebox{150mm}{!}{\includegraphics[angle=270]{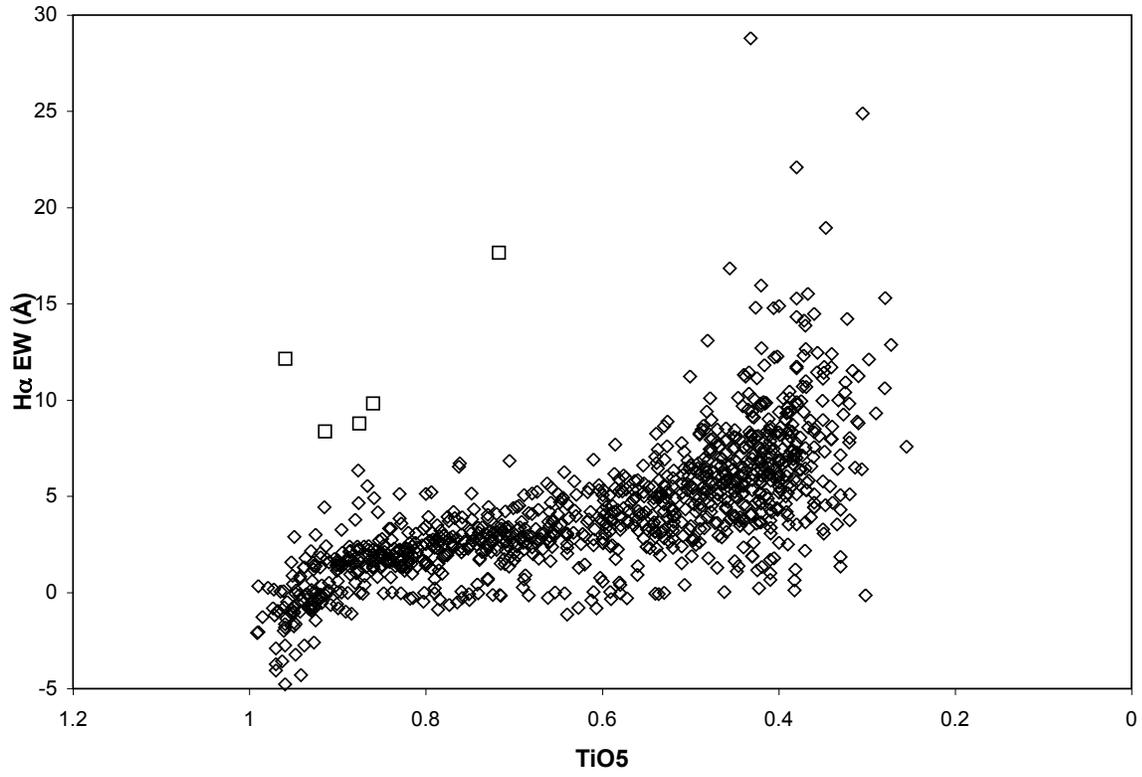}}
\caption{H$\alpha$ EW vs. the TiO5 indices.}
\end{figure}

\clearpage

\begin{figure}
\resizebox{150mm}{!}{\includegraphics[angle=270]{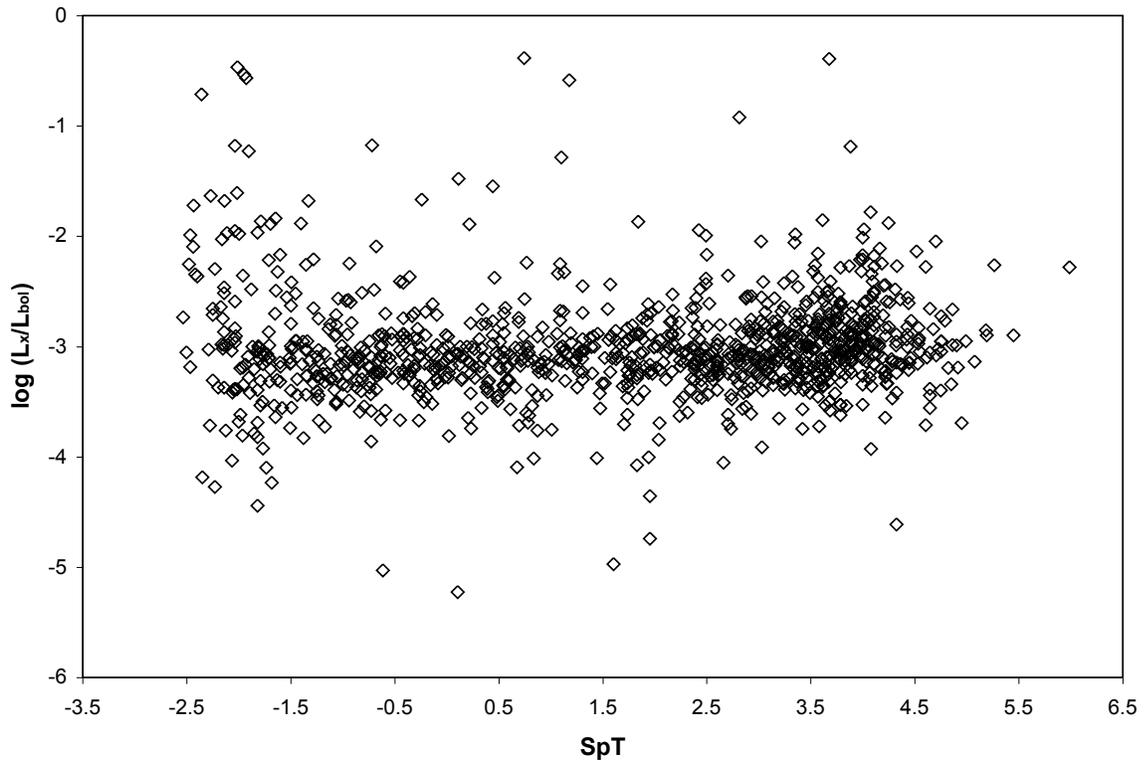}}
\caption{log $L_{X}$$\slash$$L_{bol}$ vs. SpT. Coronal emission remains saturated at a value of log $L_{X}$$\slash$$L_{bol}$ $\sim$-3, for varying SpT.}
\end{figure}

\clearpage

\begin{figure}
\resizebox{150mm}{!}{\includegraphics[angle=270]{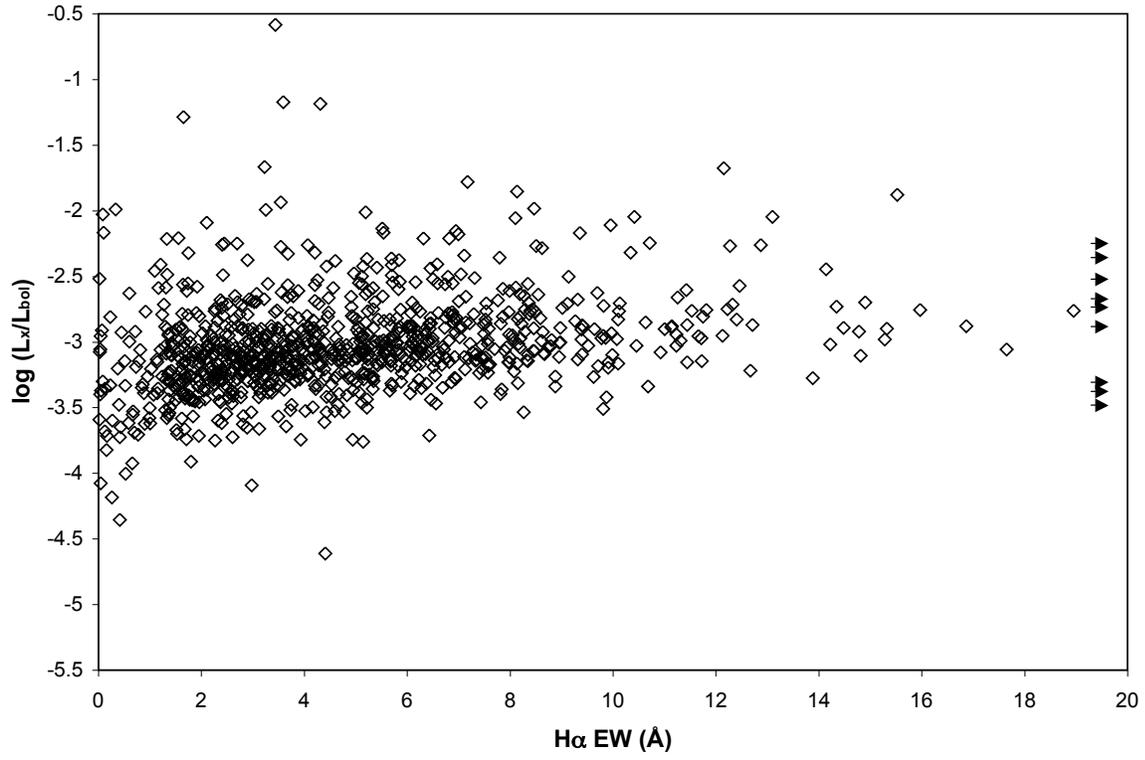}}
\caption{Saturated-type relation seen between chromospheric and coronal activity, with log $L_{X}$$\slash$$L_{bol}$ saturated at $\sim$-3, for varying H$\alpha$ EW. Arrows represent nine stars with H$\alpha$ EW $>$ 20$\AA$.}
\end{figure}

\clearpage

\begin{figure}
\resizebox{150mm}{!}{\includegraphics[angle=270]{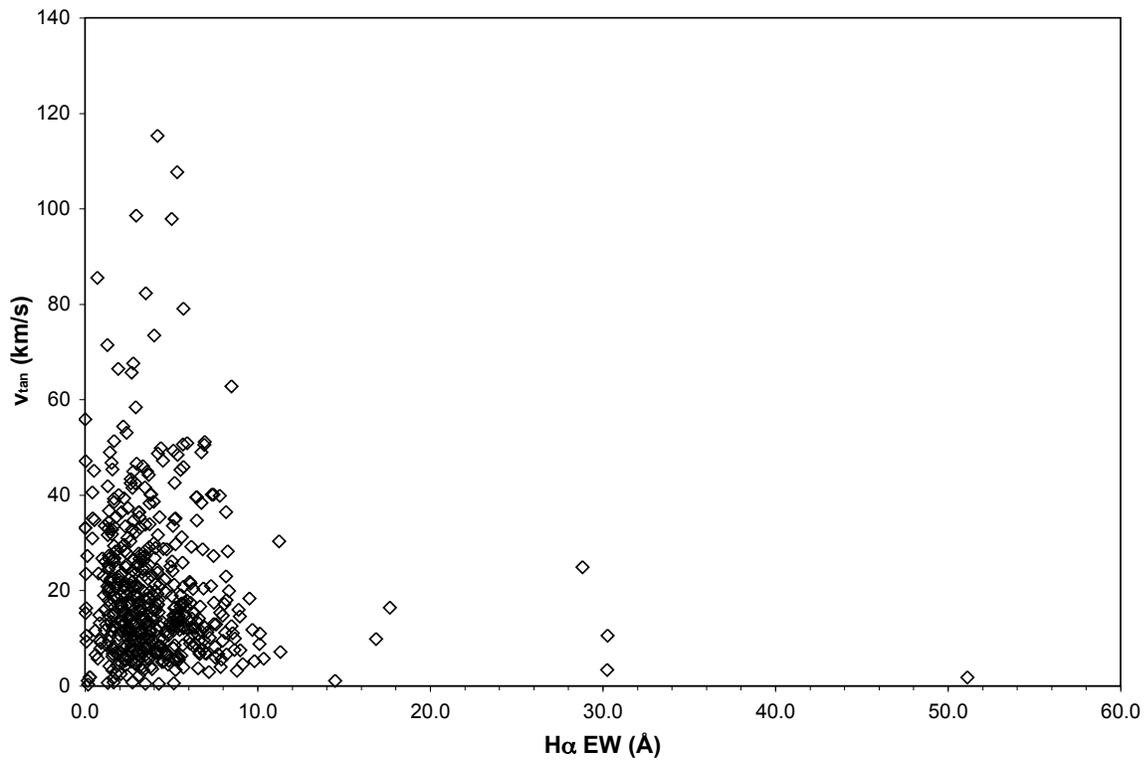}}
\caption{Tangential velocities vs. H$\alpha$ EW. Stars showing strong chromospheric activity have lower values for $v_{tan}$.}
\end{figure}

\clearpage

\begin{figure}
\resizebox{150mm}{!}{\includegraphics[angle=270]{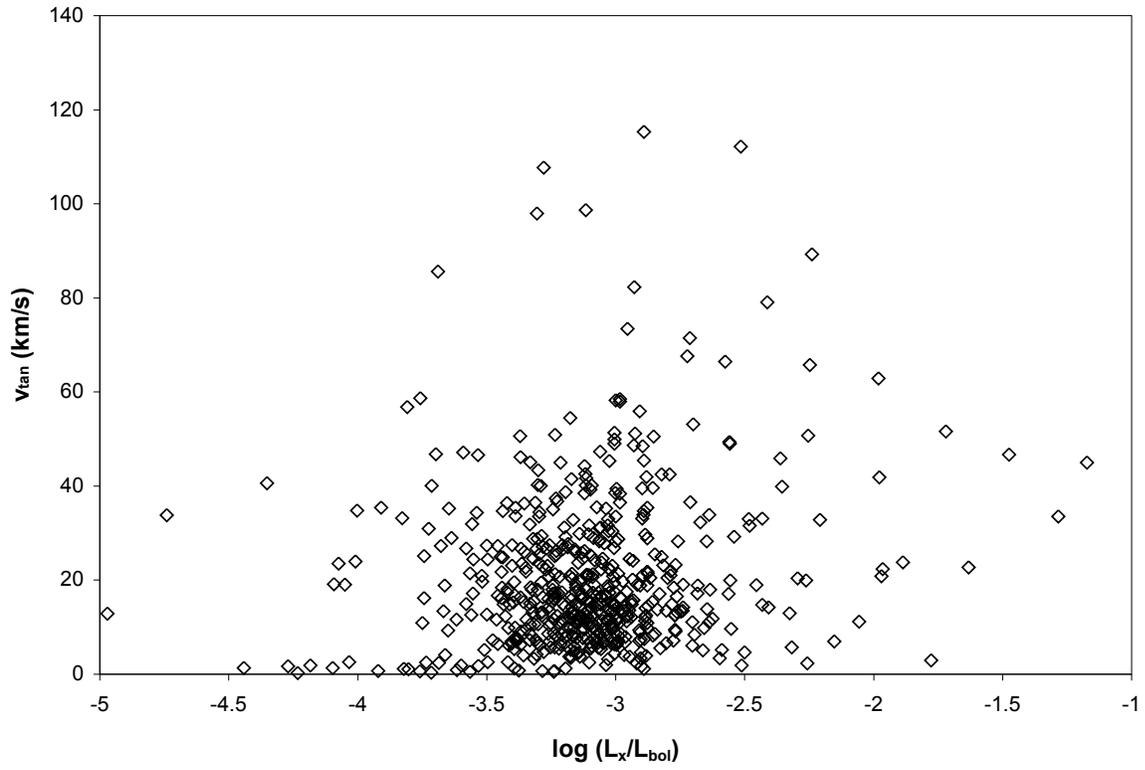}}
\caption{Tangential velocities vs. log $L_{X}$$\slash$$L_{bol}$. Coronal activity remains saturated at $\sim$-3.}
\end{figure}

\clearpage

\begin{deluxetable}{ccccccccccccccc}
\rotate
\tabletypesize{\tiny}
\tablecaption{Observed and Derived Quantities}
\tablewidth{0pt}
\tablehead{
\colhead{2MASS name} & \colhead{CaH1} & \colhead{CaH2} & \colhead{CaH3} & \colhead{TiO5} & \colhead{J} & \colhead{H} & \colhead{K}  & \colhead{SpT}  & \colhead{H$\alpha$ EW($\AA$)} & \colhead{log $L_{X}$$\slash$$L_{bol}$} & \colhead{$v_{tan}$ (km$\slash$s)} & 
\colhead{Distance (pc)} & \colhead{Distance(pc) from plx} & \colhead{Other name} 
}
\startdata
00142956+1331086 & 0.94 & 0.90 & 0.92 & 0.87 & 9.34 & 8.69 & 8.47 & K7 & 5.5 & -2.84 & 14 & 70 & -- & -- \\
0015023-725032 & 0.56 & 0.61 & 0.79 & 0.69 & 8.62 & 7.99 & 7.74 & M1 & 2.6 & -3.01 & 30 & 38 & -- & EXO 001239-7307.3 \\
00155242-2509380 & 0.98 & 0.77 & 0.84 & 0.82 & 9.48 & 8.80 & 8.63 & K7 & 1.9 & -3.08 & -- & 69 & -- & BPS CS 29503-0045 \\
00155808-1636578 & 0.94 & 0.38 & 0.60 & 0.39 & 8.74 & 8.19 & 7.91 & M4 & 5.6 & -3.08 & -- & 9 & -- & BPS CS 31060-0015 \\
00165001-0710157 & 0.82 & 0.78 & 0.86 & 0.75 & 9.58 & 8.88 & 8.69 & M0 & 4.3 & -3.09 & 13 & 64 & -- & -- \\
\enddata

\tablecomments{Table 1 is published in its entirety in the electronic edition of the {\aj}. A portion is shown here for guidance regarding its form and content.}

\end{deluxetable}

%\clearpage

\begin{deluxetable}{ccccccccccc}
\rotate
\tabletypesize{\scriptsize}
\tablecaption{Stars with strong H$\alpha$ emission}
\tablewidth{0pt}
\tablehead{
\colhead{2MASS name} & \colhead{Other name} & \colhead{TiO5} & \colhead{J} & \colhead{H} & \colhead{K} & \colhead{$v_{tan}$ (km$\slash$s)} & \colhead{H$\alpha$ EW($\AA$)} & \colhead{SpT} &
\colhead{Distance (pc)} & \colhead{log $L_{X}$$\slash$$L_{bol}$}  
}
\startdata
11015191-3442170 & TW Hya & 0.80 & 8.22 & 7.56  & 7.30 & 14  & 213.8  & M0  & 38  & -2.74 \\
11022491-7733357 & CS Cha & 0.89 & 9.11 & 8.45 & 8.20 & -- &  57.0 & K7 & 65 & -3.30\\
04374563-0119118 & EM* StHA 32 & 0.84 & 10.18 & 9.48 & 9.29 & 2 & 51.1 & K7 & 98 & -2.51\\
20194981-5816431 & -- & 0.21 & 10.66 & 10.10 & 9.72 & -- & 45.4 & M6 & 12 & -2.28\\
13220753-6938121 & -- & 0.92 & 8.28 & 7.64 & 7.29 & 11 & 30.3 & K5 & 46 & -3.31\\
05355975-0616065 & [CHS2001] 13272 & 0.85 & 10.55 & 9.89 & 9.54 & 3 & 30.3 & K7 & 118 & -2.60\\
2103599+121856 & -- & 0.43 & 11.81 & 11.19 & 10.94 & 25 & 28.8 & M3.5 & 103 & -2.82\\
05082729-2101444 & -- & 0.30 & 9.72 & 9.11 & 8.83 & -- & 24.9 & M5 & 11 & -3.19\\
07504838-2931126 & -- & 0.38 & 9.83 & 9.21 & 8.96 & -- & 22.1 & M4 & 15 &  -2.27\\
11132622-4523427 & TWA 14 & 0.72 & 9.42 & 8.73 & 8.50 & 16 & 17.7 & M0.5 & 57 & -3.06\\
13203208-7014442 & -- & 0.96 & 12.14 & 11.52 & 11.27 & -- & 12.2 & K5 & 291 & -1.68\\
18202275-1011131 & FK Ser & 0.86 & 7.64 & 6.92 & 6.62 & 5 & 9.8 & K7 & 32 & -3.51\\
06365634-0521035 & -- & 0.88 & 8.82 & 8.25 & 8.01 & 3 & 8.8 & K7 & 56 & -3.03\\
15550671-1946312 & -- & 0.91 & 9.68 & 9.05 & 8.85 & -- & 8.4 & K5 & 88 & -2.70\\
\enddata
\end{deluxetable}

%\clearpage

\begin{deluxetable}{ccccccc}
\tabletypesize{\scriptsize}
\tablecaption{Stars with $v_{tan}$ $>$ 100km$\slash$s}
\tablewidth{0pt}
\tablehead{
\colhead{2MASS name} & \colhead{Other name} & \colhead{$v_{tan}$ (km$\slash$s)} & \colhead{H$\alpha$ EW($\AA$)} & \colhead{SpT} &
\colhead{Distance (pc)} & \colhead{log $L_{X}$$\slash$$L_{bol}$}  
}
\startdata
23062530+1236570 & CCDM J23064+1236B & 152 & -0.1 & M2.5 & 101 & -1.94 \\
23154374-1221485 & L 863 -30 & 115 & 4.2 & M2 & 42 & -2.89 \\
02045317-5346162 & [FS2003] 0075 & 112 & -1.8 & K5 & 133 & -2.51 \\
16544298+4654335 & -- & 108 & 5.3 & M3 & 77 & -3.28 \\
\enddata
\end{deluxetable}

%\clearpage

\begin{deluxetable}{ccccccc}
\tabletypesize{\scriptsize}
\tablecaption{Stars within 5pc}
\tablewidth{0pt}
\tablehead{
\colhead{2MASS name} & \colhead{Other name} & \colhead{H$\alpha$ EW($\AA$)} & \colhead{SpT} &
\colhead{Distance (pc)}  & \colhead{Distance(pc) from plx} & \colhead{log $L_{X}$$\slash$$L_{bol}$}  
}
\startdata
23415498+4410407 & HH And & -0.1 & M5 & 3 & 3 & -3.69 \\
0112305-165956 & YZ Cet & 1.9 & M4.5 & 4 & 4 & -3.44 \\
0200127+130310 & TZ Ari & 1.4 & M4.5 & 4 & 4 & -3.55 \\
06045215-3433360 & AP Col & 12.1 & M5 & 4 & -- & -2.95 \\
0329197-114041 & -- & 4.4 & M4.5 & 4 & -- & -4.61 \\
\enddata

\end{deluxetable}

\end{document}